\documentclass[twocolumn,twoside,amsmath,amssymb,superscriptaddress,showpacs,nofootinbib,aps]{revtex4-1}

\usepackage{slashed}
\usepackage{amsmath}
\usepackage{amsthm}
\usepackage{subfigure}
\usepackage[utf8]{inputenc}
\usepackage{graphicx}
\usepackage{epsfig}
\usepackage{amssymb}
\usepackage{dcolumn}
\usepackage{bm}
\usepackage{color}

\newcommand{\ba}{\begin{array}}
\newcommand{\ea}{\end{array}}
\def\br{\begin{eqnarray}}
\def\er{\end{eqnarray}}
\def\be{\begin{equation}}
\def\ee{\end{equation}}
\usepackage{hyperref}

\def\({\left(}
\def\){\right)}

\def\<{\left\langle}
\def\>{\right\rangle}

\newcommand{\TQ}{\textnormal{\tiny \textsc{T(Q)}}}
\newcommand{\QT}{\textnormal{\tiny \textsc{Q(T)}}}
\newcommand{\Q}{\textnormal{\tiny \textsc{Q}}}
\newcommand{\T}{\textnormal{\tiny \textsc{T}}}
\newcommand{\E}{\textnormal{\tiny \textsc{E}}}
\def\tt{\textnormal\tiny\textsc}

\begin{document}

\title{Perturbative corrections to technicolor}

\author{A. C. Aguilar}
\email{aguilar@ifi.unicamp.br}

\affiliation{University  of  Campinas - UNICAMP,\\
Institute  of  Physics  ``Gleb  Wataghin'',  13083-859,  Campinas,  SP,  Brazil}

\author{A. Doff}
\email{agomes@utfpr.edu.br}

\affiliation{Universidade Tecnol\'ogica Federal do Paran\'a - UTFPR - DAFIS
Av Monteiro Lobato Km 04, 84016-210, Ponta Grossa, PR, Brazil}

\affiliation{Instituto de F{\'i}sica Te\'orica - UNESP, Rua Dr. Bento T. Ferraz, 271,\\ Bloco II, 01140-070, S\~ao Paulo, SP, Brazil}

\author{A. A. Natale} 
\email{natale@ift.unesp.br}

\affiliation{Instituto de F{\'i}sica Te\'orica - UNESP, Rua Dr. Bento T. Ferraz, 271,\\ Bloco II, 01140-070, S\~ao Paulo, SP, Brazil}

\begin{abstract}
The full solution of technicolor (TC) Schwinger-Dyson equations should include radiative corrections induced by extended technicolor (ETC)
(or other) interactions. We verify that when TC is embedded into a larger theory including also QCD, these radiative corrections couple
the different strongly interacting Schwinger-Dyson equations, providing a tiny mass to technifermions and changing the ultraviolet
behavior of the gap equation solution. We argue about the origin of the different quark masses without appealing for different ETC boson masses, in one scenario where most of the new physics will appear in interactions with the third fermion generation and with a TC scalar boson possibly lighter than the TC characteristic scale ($\Lambda_{\tt{TC}}$). 
\end{abstract}



\maketitle

The origin of fermion and gauge boson masses in the standard model (SM) of elementary particles is explained by
their interaction with the Higgs boson. The discovery of this boson at the LHC~\cite{atlas,cms} has crowned the SM; however,
the data still cannot discard the possibility of this boson being a composite one. 
The case of a composite state, generating dynamical gauge symmetry breaking,  instead of an elementary one is more akin to the phenomenon of spontaneous symmetry breaking that originated from the Ginzburg-Landau Lagrangian. The latter can be derived from the microscopic BCS theory of superconductivity describing the electron-hole interaction, which can be interpreted as a composite state. A similar mechanism
happens in QCD where the chiral symmetry breaking is promoted by a nontrivial vacuum expectation value of a fermion
bilinear operator and the Higgs role is played by the composite $\sigma$ meson. In particular, the technicolor TC idea was
the earliest attempt to build models in this direction~\cite{wei,sus}.

The main ideas about TC models were reviewed in Refs.~\cite{far,mira} and recent phenomenological studies about this class of
models can be seen in Refs.~\cite{an,sa,sa1,sa2,sa3,sa4,be} and references therein. Despite the fact that TC models are much more complex than the ones with elementary scalar bosons, the main difficulty to build a realistic  model lies in the ordinary
behavior of the technifermion self-energy that is proportional to 
$\Sigma_{\T} (p^2) \propto \frac{\mu_{\tt{TC}}^3}{p^2} (p/\mu_{\tt{TC}})^{\gamma}$
where $\mu_{\tt{TC}}$ is the characteristic TC dynamical mass at zero momentum and $\gamma$ the anomalous mass dimension. This self-energy leads to the known quark mass ($m_{\Q}$) given by $m_{\Q} \propto \mu_{\tt{TC}}^3/M_{\E}^2$, where $M_{\E}$ is the mass of an extended technicolor boson (ETC), which is a particle that may change flavors. In order to describe, for example, the top quark mass we need a small $M_{\E}$ value, and this boson generates flavor changing neutral currents at one undesirable level. A possible solution to this dilemma requires a large $\gamma$ value~\cite{holdom}, which can be obtained either with the introduction of (i) a large number of fermions or (ii) with a gauged four-fermion interaction~\cite{lane0,appel,yamawaki,aoki,appelquist,shro,kura,yama1,yama2,mira2,yama3,mira3,yama4}. Regardless of  all these efforts, in these dynamical symmetry breaking models, it has not been clear up to now why the heaviest quark has a current mass of {\cal{O}}$(100)$ GeV  whereas the light quarks have a current mass of few MeV.  In addition, in the context of dynamical gauge symmetry breaking models, it is not naturally expected to have a scalar boson (the Higgs boson in this case) with a mass smaller than the Fermi or TC scale.
\begin{figure}[h]
\centering
\includegraphics[scale=0.45]{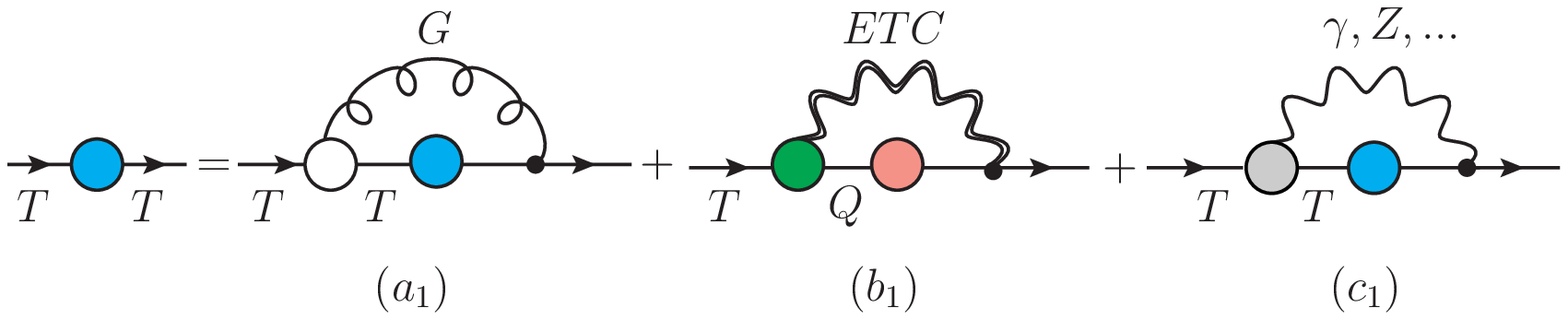} 

\vspace{0.25cm} 

\includegraphics[scale=0.45]{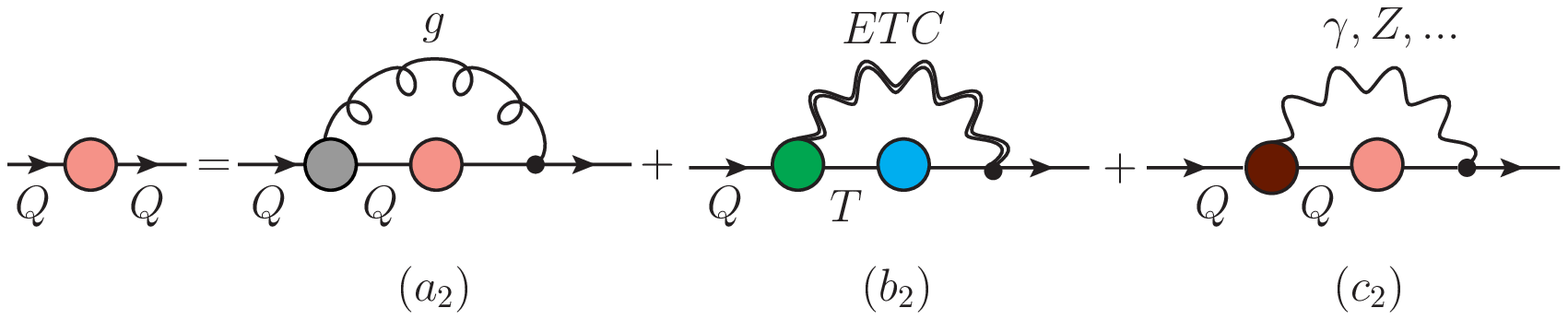}
\vspace{-0.25cm} 
\caption[dummy0]{The coupled system of  SDEs for TC  (T$\equiv$technifermion) and QCD  (Q$\equiv$quark)  including  ETC and electroweak or other corrections. $G \,(g)$ indicate technigluons (gluons).}
\label{fig1}
\end{figure}

In this work, we argue that the answer to these questions may come out when QCD and TC are embedded into a larger group, possibly ETC or a grand unified theory (GUT)~\footnote{One example of a model where TC and QCD are embedded into a larger ETC theory is the Farhi-Susskind GUT model~\cite{suss}.}. However there are two crucial requirements for this  scenario  to work out:  (i) the interconnection of the SDEs describing the TC and QCD self-energies, as shown in 
Fig.~\ref{fig1}, and (ii)  the inclusion of 
a horizontal or family symmetry, where the third quark generation couples preferentially to TC and the first quark generation to QCD.  As we see, a direct consequence of point (i) is 
the generation a hard TC self-energy which allows  for a scalar boson mass smaller than $\Lambda_{\tt{TC}}$\footnote{Note that the ETC group can be composed either by TC and QCD as 
in the Farhi-Susskind model or it may include the electroweak group as well. As it will
be discussed later, the ETC role could be even played by a GUT that should include TC, QCD and the electroweak theory. This is going to be possible because the fermion masses in our
scenario will be weakly dependent on the GUT mass.}.

The starting point in this analysis is the diagrammatic representation of the coupled SDEs for the techniquark and quark  self-energies,  shown in the first 
and second lines of Fig.~\ref{fig1}, respectively. In this figure, the curly lines correspond to gluons (g) or technigluons (G)  and the wavy lines to the ETC (double) or other weakly interacting bosons\footnote{The SDE of the Fig.~\ref{fig1} is similar to the ones describing the photon perturbative corrections to the quark mass~\cite{brodsky}, where it is understood that the strong interactions should be summed first~\cite{craigie}.}.

Notice that the above coupled system is rather intricate. More specifically, it involves different full boson propagators and fully dressed vertices which should be closely intertwined through the different mass scales of the
theories, namely  ($\Lambda_{\tt{QCD},\tt{TC},\tt{ETC}}$). Here, we restrict ourselves to exploring the result of this coupled system in a rather simplified
context. First, without specifying a model, we assume that QCD and TC are embedded into a gauge group, like the $SU(5)_{\tt{S}}$ of the Farhi-Susskind model. Then, we neglect the
possible contributions that the diagrams $(c_1)$ and $(c_2)$ may give and later discuss their effects. Therefore, in our 
analysis, the conventional self-energies of the techniquarks $(a_1)$ and quarks $(a_2)$ only receive the  perturbative corrections generated by the  ETC interaction which couples 
 the techniquarks to quarks (and vice versa), as  represented by the diagrams $(b_1)$ and $(b_2)$. Finally,  we approximate  the  fully dressed propagators and vertices, entering into the gap equations, by  their tree level expressions, ignoring the fact that
the gauge bosons of the two strongly interacting theories are dynamically massive as suggested in a series of works~\cite{cornwall,Aguilar:2008xm,Aguilar:2010cn,us}.

After applying the above considerations, we arrive in the following coupled system
of integral equations  for the self-energies of the techniquarks
and quarks, respectively (in Euclidean space) 
\begin{eqnarray}
\Sigma_{\T}(p^2)&=&3\lambda_{\T}\!\int_{k} \frac{\Sigma_{\T}(k^2)}{(p-k)^2[k^2+\Sigma^2_{\T}(k^2)]} + m_{\T}(p^2), \label{eq1} \\
\Sigma_{\Q}(p^2)&=&3\lambda_{\Q}\!\int_{k}\frac{\Sigma_{\Q}(k^2)}{(p-k)^2[k^2+\Sigma^2_{\Q}(k^2)]}+ m_{\Q}(p^2),\label{eq2}
\end{eqnarray}
with
\be
m_{\TQ} (p^2) =3\lambda_{\E}\!\int_{k} \frac{\Sigma_{\Q(\T)}(k^2)}{[(p-k)^2 + M^2_{\E}][k^2+\Sigma^2_{\Q(\T)}(k^2)]}\,,
\label{eq31}
\ee
where we have introduced the compact notation \mbox{$\int_k = {1}/({2\pi^2})\int^{\infty}_0 dk^2 k^2\int^{\pi}_{0} d\theta\sin^2\theta$} with $\theta$ being 
the angle between the momenta $p$ and $k$. Moreover, we defined $\lambda_i=C_i\alpha_i$ with $\mbox{i=\mbox{T, Q}}$ and $\mbox{E}$, where $\alpha_i$ are the TC, QCD and ETC coupling constants, respectively. The  $C_i$ are the corresponding Casimir eigenvalues for the different fermionic representations, {\it i.e.} $C_{\Q}=4/3$,  $C_{\T}=3/4$ and $C_{\E}=1$, where this last value was chosen
for simplicity, since we are not going to define one specific model for the ETC interaction. 
 
We can easily identify the second term of Eq.~(\ref{eq2}) as an~\emph{effective current quark mass} obtained through TC interaction, described by diagram ($b_2$). With the
appropriate values for $\lambda_{\Q}$, $\lambda_{\E}$, and $M_{\E}$, we obtain a solution that is the sum of the dynamical
quark mass with its effective current mass. In a similar way, Eq.~(\ref{eq1}) provides the dynamical techniquark mass with a very tiny effective current mass generated
by the QCD correction, represented in the diagram ($b_1$). If we perform a four-fermion approximation for the ETC contributions, by taking the limit of large $ M_{\E}$  of the diagrams ($b_1$) and ($b_2$), the effective current masses added to the SDEs reduce to
\be
m_{\TQ} \propto \frac{\lambda_{\E}}{4\pi M^2_{\E}}\int^{M^2_{\E}}_0\!\!\!\! dk^2 \,\Sigma_{\QT}(k^2) \, .  
\label{eq2a}
\ee
With the approximation performed in Eq.~(\ref{eq2a}), which is equivalent to adding a bare mass, the solutions of Eqs.~(\ref{eq1}) and (\ref{eq2}) are a superposition of the regular [$\propto 1/p^2$] plus irregular  [$\propto \ln (p^2)^{-\gamma}]$
solutions~\cite{lane2,lang,pag}.  Nowadays, it is known that the  SDE solutions 
may vary between these two behaviors according
to the boundary conditions~\cite{holdom,lane0,appel,yamawaki,aoki,appelquist,yama1,yama2,mira2,yama3,mira3,yama4,cg,takeuchi,us1}, but
we certainly can expect that a slowly decreasing with momentum self-energy dominates the large $p^2$ behavior. 

We have solved numerically the coupled Eqs.~(\ref{eq1}) and~(\ref{eq2}) using $\alpha_{\T} = 8.0$ and $\alpha_{\Q} =0.87$. The coupling values were chosen such that in isolation  (i.e. $\alpha_{\E}=0$)  the dynamical techniquark and quark masses generated are respectively \mbox{$\mu_{\tt{TC}}\approx 1$ TeV} and $\mu_{\tt{QCD}} \approx 0.3$ GeV, and
the solutions decrease as $1/p^2$. When the ETC interaction is turned on, assuming $\alpha_{\E}=0.032$ and \mbox{$M_{\E} = 100$ TeV}, which was assumed as a typical
ETC mass scale~\cite{far,suss}, we verified that the solutions seem to decrease
like $1/p^2$ after the $\mu_{\tt{TC}(\tt{QCD})}$ scale and appear to be basically flat at large momenta, consistent with a superposition of
the regular and irregular solutions, although these large momenta were of the order of our numerical cutoff. This behavior is not unexpected, since the quark condensation gives a tiny mass to technifermions, however the asymptotic
behavior, due to the current mass, is overwhelmed by the large dynamical TC mass, and it is difficult to extract a clear signal of the
superposition of the different solutions from the full $\Sigma_{\T}(p^2)$ behavior. 

Another way to verify that the self-energies decrease slowly with the momenta is to determine their anomalous dimension. As already mentioned, much of the information about chiral symmetry breaking resides in the boundary conditions of the SDE gap equation~\cite{cg}, from where we can derive
the anomalous dimension, as shown in Ref.~\cite{us1}. Using this observation, we determined an effective four-fermion coupling constant 
\mbox{$\kappa_{\E}  \propto  C_{\E}\alpha_{\E}f(m_{\T}(M_{\E}),m_{\Q}(M_{\E}))$}~\cite{us2}, and when
this constant $\kappa_{\E}$ is introduced into Eq.~(15) of Ref.~\cite{us1}, using $\mu_{\tt{TC(QCD)}}$ as the dynamical $\tt{TC}(\tt{QCD})$ masses of our coupled equations, we obtain $\gamma_{\T} \sim 2$. This indicates again a hard asymptotic behavior for  $\Sigma_{\T}(p^2)$, corroborating the fact
that the self-energies are changed when we consider the radiative corrections for the SDE, however, due to the approximations of Ref.~\cite{us1}, it is not possible to verify 
with high precision how the asymptotic behavior of $\Sigma_{\T}(p^2)$ and $\Sigma_{\Q}(p^2)$ are modified in the coupled system.

The best way to verify how the asymptotic behavior of $\Sigma_{\T} (p^2)$ [or $\Sigma_{\Q} (p^2)$] has changed to an irregular-type solution
is to compute the quark masses as a function of the ETC mass. This behavior is extremely dependent on the asymptotic self-energy.
To observe this, let us suppose that the TC self-energy is
\be
\Sigma_{\T}(p^2)\approx \mu_{\tt{TC}} \left( \frac{\mu_{\tt{TC}}^2}{p^2+\mu_{\tt{TC}}^2}\right),
\label{eqx1}   
\ee  
which was conveniently normalized to the dynamical techniquark mass $\mu_{\tt{TC}}$ as $p^2\rightarrow 0$ and decays asymptotically as $1/p^2$. Substituting 
Eq.~(\ref{eqx1}) into Eq.~(\ref{eq31}), we obtain, in the limit of  zero momentum, that  
\be
m_{\Q} \propto \lambda_E \frac{\mu_{\tt{TC}}^3}{M^2_{\E}} \,.
\label{eqx2}
\ee
On the other hand,  if we assume that the TC self-energy is giving by an irregular-type solution,
which can be cast in the form 
\be
\Sigma_{\T}(p^2)\approx \mu_{\tt{TC}} \left[ 1+ \delta_1 \alpha_T \ln\left[(p^2+\mu^2_{\tt{TC}})/\mu^2_{\tt{TC}}\right] \right]^{-\delta_2} \,,
\label{eqx3}
\ee
where $\delta_1$ and $\delta_2$ are constants depending on the TC gauge group and fermionic representation. Note that  Eq.~(\ref{eqx3}) is also normalized to $\mu_{\tt{TC}}$ as $p^2\rightarrow 0$. It follows that the resulting quark mass calculated from Eq.~(\ref{eq31}) in this case is given by~\cite{nat}
\be
m_{\Q} \propto \lambda_E \mu_{\tt{TC}} [1+\delta_1 \alpha_T \ln(M^2_{\E}/\mu_{\tt{TC}}^2)]^{-\delta_2} \, .
\label{eqx4}
\ee
These quite different behaviors that may result for quark masses are one clear identifier of the asymptotic behavior of $\Sigma_{\T}(p^2)$ [and the same can be formulated with respect to $\Sigma_{\Q}(p^2)$]. We stress that $m_{\Q}$ is the current mass at \textit{zero momentum}, and the total mass
should also include the dynamical mass and all momentum dependence.

\begin{figure}[h]
\centering
\includegraphics[width=1.00\columnwidth]{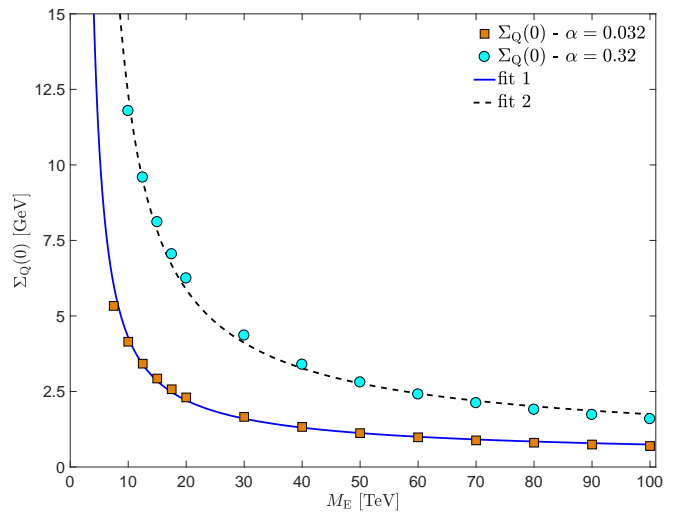}
\caption[dummy0]{The comparison of the behavior of
$\Sigma_{\Q} (0)$ as a function of $M_{\E}$ 
with the fit given by Eq.~(\ref{eqx5}). Fit $1$ was obtained with
\mbox{$\alpha_{\E}= 0.032$} and fit $2$  with \mbox{$\alpha_{\E}= 0.32$}; the other parameters are described in the text.}
\label{fig2}
\end{figure}

Turning on the ETC interactions (by choosing either \mbox{$\alpha_{\E}=0.032$} or \mbox{$\alpha_{\E}=0.32$}), we study numerically Eqs.~(\ref{eq1}) and~(\ref{eq2}) and determine how 
$\Sigma_{\Q}(0)$ behaves with $M_{\E}$ for the coupled SDE system. This behavior can be seen in Fig.~\ref{fig2}. Because of the fact that, the dynamical
quark mass is negligible in comparison with the current mass, it  turns out that $ \Sigma_{\Q} (0) \cong m_{\Q} $. Then, $\Sigma_{\Q}(0)$ 
can be accurately fitted by 
\be
m_{\Q}^{fit} = a_i [\ln (M^2_{\E}/\mu_{\tt{TC}}^2)]^{-b_i} \, ,
\label{eqx5}
\ee
which is precisely the same behavior found in Eq.~(\ref{eqx4}). 
More specifically, the set of optimal values when \mbox{$\alpha_{\E}=0.032$} is  
\mbox{$a_1= 203.92$ GeV} and \mbox{$b_1= 2.53$}, 
to be denoted by fit $1$. For fit $2$, defined  when \mbox{$\alpha_{\E} = 0.32$},  we have \mbox{$a_2= 912.9$ GeV} and \mbox{$b_2= 2.82$}. The reduced chi square of both fits is \mbox{$R^2=0.99$}. The quark mass turns out to be heavy and of order of a few GeV as can be seen in Fig.~2. However, we have to
remember that in the numerical calculation, we used $C_{\E} = 1$, which is a rather naive assumption, since any theory embedding TC and QCD is represented by a large gauge group with fermions in higher dimensional representations and increased $C_{\E}$ values. 
Moreover, as in the GUT model of Ref.~\cite{suss}, one specific quark may obtain mass from different diagrams classified as $(b_2)$ 
in Fig.~\ref{fig1}, which when added generate  
masses up to {\cal{O}}($100$) GeV. These masses are roughly
proportional to $\mu_{\tt{TC}}$ [of {\cal{O}}($1$) TeV] times $\lambda_E$. 

The fit of Eq.~(\ref{eqx5}) shows unmistakably  how the logarithmic factor enters
into action in Fig.~2, and is consistent with the prediction of Eq.~(\ref{eqx4}). It must be remembered that $m_{\Q}$, given by Eq.~(\ref{eq31}), also runs and the full result of the quark mass should include the momentum dependence. The large coefficients $a_i$ appearing
in Eq.~(\ref{eqx5}) are just a consequence of extending our fit, which is only appropriate  for the asymptotic regime of the dynamical masses,  
to small $M_{\E}$ values.  We recall that  such small $M_{\E}$ values are not phenomenologically acceptable
mass scales, if this theory is supposed to embed TC and QCD. We also do not expect that technigluon(gluon) masses and vertex corrections affect these
results. Notice that in this scenario it is perfectly possible to obtain an infrared mass of 
{\cal{O}}$(100)$ GeV, which is of the order of the top quark mass. 
 
The coupled SDE system indicates that the techniquarks obtain a dynamical mass ($\mu_{\tt{TC}}$) that at zero momentum is of order of $1$ TeV. The self-energy momentum
dependence is a superposition of the regular and irregular solution, whose asymptotic behavior is dominated by the irregular
solution that appears due to a tiny current mass generated by the QCD condensation. Quark masses also have a dynamical
mass of the order of $300$ MeV; however, they can obtain a current mass from TC condensation up to {\cal{O}}$(100)$ GeV, which can explain the
third generation quark masses. In this case, the total quark mass is totally dominated by the irregular solution, i.e. the one that
runs with the momentum as logarithm. Apart from the small logarithmic dependence on $M_{\E}$, the ordinary quark masses are always proportional to $\alpha_{\E} C_{\E} \mu_{\tt{TC}}$, as given by Eq.~(\ref{eqx4}). The technifermion masses will be mostly dynamical and proportional to $\mu_{\tt{TC}}$. However this scenario is not complete until we take into account all radiative corrections.

We have neglected the diagrams $(c_1)$ and $(c_2)$ of Fig.~1. All self-energies in these diagrams also run logarithmically
with the momentum, and the diagram $(c_1)$ generates for the techniquarks an effective mass proportional to $\mu_{\tt{TC}}$ times, for
example, an electroweak charge or any other
charge of interaction that contributes to the third diagram of Fig.~1, and, depending on the model, an ETC charge. Ordinary quarks also obtain a mass of a few MeV, which may appear, due to the electroweak or other interactions, depicted in the diagram $(c_2)$. For instance, for the QED
interaction we have $m_{\Q} \propto\alpha_{em} e_q^2 \mu_{\tt{QCD}}$, 
where $\alpha_{em}$ is the QED coupling constant and $e_q$ the quark charge. This fact leads to a quite interesting solution for the fermionic mass spectra.
In this scenario we must have a family or horizontal symmetry imposing that the third quark family couples preferentially to 
TC,  i.e. the diagram  $(b_2)$ of Fig.~1, whereas the first quark family obtains masses 
preferentially from QCD, receiving most of the radiative contributions from the diagram $(c_2)$. The final quark mass
matrix can be of the Fritzch type
\br
 m_f =\left(\begin{array}{ccc} 0 & A & 0\\ A^* & 0 & B \\
0 & B^* & C
\end{array}\right),
\label{ze13} 
\er
where $A\approx \alpha_{em} e_q^2 \mu_{QCD}$ and $C\approx \alpha_{\E} C_{\E} \mu_{\tt{TC}}$, providing a natural explanation of the 
different mass scales. The factor $B$ should also appear naturally and be between $A$ and $C$, because the TC and QCD scalars mix, due to the many interactions that may connect QCD and TC.

To produce a mass matrix like the one of Eq.~(\ref{ze13}) we can choose as a horizontal symmetry the  
$SU(3)_{\tt{H}}$ group assigning to the first and third family different quantum numbers, in such a way that the third family couples only to TC and the first one only to QCD. Higher order
loop contributions to the SDE  give intermediate masses to the second quark generation, which explains the origin of the term $B$. 

To verify the origin of the term B, let us represent the TC and QCD scalar composite fields by $\eta$ and $\phi$ that will be
formed in  the ${\bf \overline{6}}$ and  ${\bf 3}$ representations of $SU(3)_{\tt{H}}$. The most
general effective potential is described by  
\be
 V(\eta,\phi) = \mu^2_{\eta}\eta^{\dagger}\eta + \lambda_{\eta}(\eta^{\dagger}\eta)^2 +
\mu^2_{\phi}\phi^{\dagger}\phi + \lambda_{\phi}(\phi^{\dagger}\phi)^2,
\label{vh}
\ee
where we can identify the vacuum expectation values (vevs) of the TC and QCD condensates as given by the ratio of their respective masses and couplings, {\it i.e}
\be
v^{2}_{\eta}=-\frac{\mu^2_{\eta}}{\lambda_{\eta}}\,\,\,,\,\,\,v^{2}_{\phi}
=-\frac{\mu^2_{\phi}}{\lambda_{\phi}} ,
\label{mc}
\ee

Such potential is quite plausible if we consider the results of Refs.~\cite{Carpenter:1988qu,Carpenter:1989ij}, where it was
shown that the interactions of a composite Higgs boson are very similar to the ones of a fundamental boson.
This system leads to an intermediate mass scale and to a mass matrix identical
to Eq.~\eqref{ze13}.

The QCD and TC vevs, due to the horizontal symmetry, can be written
respectively in the following form~\cite{Wilczek:1978xi,Gelmini:1983cd}
\be
 \langle\eta\rangle \sim 
\left(\begin{array}{c} 0 \\ 0 \\  v_{\eta}
\end{array}\right)\,\, \,,\,\,\, 
\langle\phi\rangle \sim 
\left(\begin{array}{ccc} 0 & 0 & 0\\ 
0 & 0 & 0 \\ 
0 & 0& v_{\phi} 
\end{array}\right),
\label{veta}
\ee
which are of the order of approximately  $250$ MeV and $250$ GeV.
We can now verify what fermionic mass matrix one can obtain with the
vevs of Eq.~(\ref{veta}). Assuming that the composite scalars $\eta$ and
$\phi$ have ordinary Yukawa couplings to fermions described by the
following effective Yukawa Lagrangian
\be 
{\cal{L}}_{Y} = a\bar{\Psi}^{i}_{L\lambda}\eta^{k}_{\lambda}U^{j}_{R}\epsilon_{ijk} +
b\bar{\Psi}^{i}_{L\lambda}\phi^{ij}U^{j}_{R} ,
\label{y1}
\ee
where  $\Psi$ and $U$ are the ordinary fermion fields. In addition, $\lambda$ is a weak hypercharge $SU(2)_{w}$ index. For instance, $\lambda = 1$ represents charge $2/3$ quarks and $\lambda  = 2$ correspond to the charge $1/3$
quarks. In addition, $i,j$ and, $k$ indicate the components of the composite scalar bosons of
the representations ${\bf 3}$ and $\overline{\bf {6}}$ of $SU(3)_{H}$; $a$ and $b$ are the coupling constants.
Substituting the vevs of Eq.~(\ref{veta}) in the Yukawa Lagrangian for the charge $2/3$ quarks, we obtain
\be {\cal{L}}_{Y} = a\bar{c}_{L}v_{\eta}u_{R} - a\bar{u}_{L}v_\eta{c}_{R} + b\bar{t}_{L}v_{\varphi}t_{R} ,
\label{yukl}
\ee
leading to a mass matrix in the $(u\,,c\,,t)$ basis that is given by
\be 
\overline{m}^{\frac{2}{3}} = \left(\begin{array}{ccc} 0 & -av_{\eta} & 0\\ av_{\eta} & 0 & 0 \\ 0 & 0 & bv_{\phi}
\end{array}\right). 
\ee
The third generation fermions obtain large masses because coupling
directly to technifermions, while the first generation ones obtain masses coupling to ordinary quarks.  Having this picture in mind, we can now see that the most general vev for this
system includes the mass generation for the intermediate family.

\begin{figure}[ht]
\begin{center}
\includegraphics[width=1.00\columnwidth]{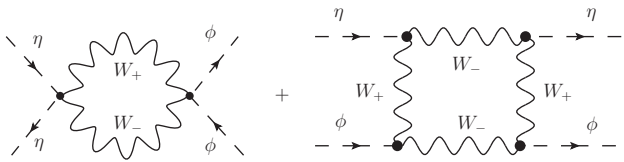} 
\caption{Higher order corrections coupling the $\eta$ and
$\phi$ composite bosons. The effective coupling between two scalars and two W's occurs through a fermion loop. } 
\label{f3}
\end{center}
\end{figure}

Note that there is no way to prevent the coupling at higher order of
the different composite scalar bosons with $SU(3)_H$ quantum numbers. Examples of such couplings are shown in Fig.~\ref{f3}, where
the effective coupling between scalars and gauge bosons involves the self-energy solution that we have discussed, and is
also enhanced due to its hard behavior with the momentum.

\par These diagrams produce the following contribution to the potential 
\be V_{\delta}(\eta,\phi) =  \lambda_1 \eta^{\dagger}\eta\phi^{\dagger}\phi +
\lambda_2 \eta^{\dagger}\phi\eta\phi^{\dagger} + ... \label{vh2} \ee
which should be added to Eq.~(\ref{vh}), shifting the vev matrix in order
to reproduce a fermionic Fritzch mass matrix of Eq.(10). Details about this mechanism can be seen in Refs.~\cite{Wilczek:1978xi,Gelmini:1983cd,us2003}.

 Therefore, the mass generation
of the different fermion families is just one effect of the alignment of the two different strong interactions in the presence of
the horizontal symmetry. Another type of model, which naturally suppresses the TC coupling to the light quarks, can be formulated
choosing the technifermion representation under QCD in such a way that they do not couple, at leading order, to the first two quark generations.
 
Up to now we have not discussed the leptonic mass spectra. We 
say that the family or horizontal symmetry should also be extended to the leptonic sector. In this
case only the $\tau$ lepton obtains mass from technifermion condensation, and the other first two leptonic generations remain massless. More
complicated mass spectra  depend on specific models. In order to give an idea of the possible different models, let us now indicate techniquarks and technileptons 
by T and L, whereas ordinary quarks and leptons are indicated by $Q$ and $\ell$, respectively, and assume that there is a family symmetry imposing that the
TC sector connects only to the third ordinary fermion generation. If technileptons couple only to themselves and to ordinary leptons (i.e. there is not
LT or LQ coupling) the technilepton self-energy is decoupled from the others, and its solution is soft ($\propto 1/p^2$). In this case 
only the $\tau$ lepton obtains a tiny mass.  If  we now admit that the  theory has LT and/or LQ couplings, then we have a new set of coupled equations. Now the technilepton self-energy is hard, 
as in the techniquark case, and the $\tau$ lepton may obtain a few GeV mass. Finally, a leptonic mass matrix like the one of Eq.~(\ref{ze13}), will be generated only at higher
perturbative level, when ordinary quarks and leptons may be unified at a deeper level (with, for instance, $\ell$Q couplings), and probably related to some very heavy unified 
gauge boson mass. Generating naturally $\ell$ masses smaller than Q masses.  Note that the larger gauge group is unifying these theories the more complex the
set of coupled SDE. Of course, here we are assuming only left-handed neutrinos, and a full explanation of the fermionic mass 
spectra (including neutrino masses) is a hard task and out of the scope of this work.

It is interesting to give an idea of what kind
of model it is possible to build in this scenario. We can follow similar ideas as the ones in the Farhi-Susskind model. In that model, the $SU(5)_{\tt{S}}$ group was broken to $SU(2)_{\tt{HC}}$ and $SU(3)_c$, where $HC$ indicates the hypercolor (or TC) theory, however it was not discussed in detail how the
$SU(2)$ scale $\Lambda_{\tt{HC}}$ could be larger than $\Lambda_{\tt{QCD}}$. In principle, we can adopt the most 
attractive channel hypothesis to have one idea about the symmetry breaking of the larger group. It is not impossible 
that the symmetry breaking pattern of a quite large group leads to a group smaller than $SU(3)_c$ with a fundamental scale
larger than $\Lambda_{\tt{QCD}}$, since this is a quite model dependent problem. However, the most probable breaking of a larger group into QCD plus another stronger interaction with $\Lambda_{\tt{TC}}>\Lambda_{\tt{QCD}}$ would happen when TC is given at least by a $SU(4)_{\tt{TC}}$ group.

 Now, let us consider a unified theory based on the
$SU(9)$ gauge group, containing a $SU(4)_{\tt{TC}}$ TC theory and the standard model. The anomaly free fermionic representations
of this theory are~\cite{Frampton:1979gt}
\be 5\otimes[9,8]_i \oplus 1\otimes [9,2]_i\,, 
\ee

\noindent where $i=1,2,3$ is a family or horizontal index that appears due to the necessary
replication of families associated to a $SU(3)_{\tt{H}}$ horizontal group; the $[\underline{8}]$ and $[\underline{2}]$ are \,antisymmetric \,under \,$SU(9)$.
These representations\, can\, be\, decomposed \,according \,to\, the\,
\,group \,product \,\,$SU(4)_{\tt{TC}}\otimes SU(5)_{gg}$, where $SU(5)_{gg}$\, is \,the\,
\,standard\, Georgi-Glashow \,GUT~\cite{Georgi:1974sy}. The  technifermions in this model transforming as $[4,5]_i$ and $[\bar{4},1]_i$ should
have different quantum $SU(3)_{\tt{H}}$ numbers than the ordinary fermions transforming as $[1,10]_i$ and $[\bar{6},1]_i$ in order
to produce a matrix like Eq.(10). 
According to the most attractive channel hypothesis~\cite{Cornwall:1974hz,Raby:1979my}, for the TC and QCD condensates (and their scalar bosons) appearing
respectively in the ${\bf \overline{6}}$ and  ${\bf 3}$ of the $SU(3)_{\tt{H}}$, as discussed previously, it is enough that the standard
left-handed (right-handed) fermions transform as triplets (antitriplets) under $SU(3)_{\tt{H}}$.
Evidently,  the full set of coupled SDE in this specific model is extremely complex involving
couplings of the $SU(9)$, $SU(5)_{gg}$, SM ones, and the horizontal bosons, and all their implications are analyzed in a
future work.

We may wonder what happens with pseudo-Goldstone bosons in the scenario we are proposing here. Diagram $(a_1)$ of Fig.~\ref{fig1} generates a dynamical TC mass at a TeV scale. However, assuming that technifermions have an electroweak or other similar charge, the diagrams $(b_1)$ and $(c_1)$ of Fig.~\ref{fig1} generate effective ``bare"
masses \mbox{$m_T \propto \alpha_i \mu_{TC} \propto O(1-10)$\,GeV}. In this case, we can obtain a lower bound on the pseudo-Goldstone masses ($m_\Pi$) using the Gell-Mann-Oakes-Renner relation
to estimate \mbox{$m_\Pi^2 \approx m_T {\<{ \bar{\psi}_T}\psi_T\>}/2F_{\Pi}^2$}, where in the right-hand side we have the TC condensate divided by the technipion decay constant. This
relation roughly implies a lower bound of \mbox{$O(30-90)$\,GeV} for the pseudo-Goldstone masses. 

\begin{figure}[h]
\centering
\includegraphics[width=0.7\columnwidth]{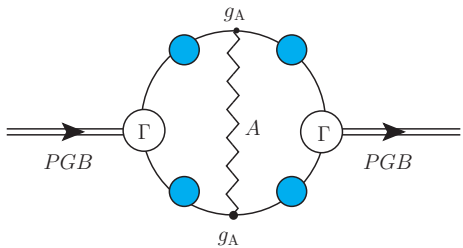}
\caption[dummy0]{Radiative correction to the mass of the pseudo-Goldstone
boson.}
\label{fig3}
\end{figure}

Another way to observe
the increase of the pseudo-Goldstone masses is to compute the effect of electroweak (or other) radiative corrections to these bosons. The diagram of Fig.~\ref{fig3} shows the radiative correction to the pseudo-Goldstone boson mass induced by one gauge boson A with coupling constant 
$g_A$, mass $M_A$, and a vertex indicated by $\Gamma$ that is proportional to the technipion wave function.
It is quite important to remember that the technipion wave function is related to the technifermion self-energy as 
\mbox{$\left.\Phi_{BS}^\Pi (p,q)\right|_{q\rightarrow 0} \approx \Sigma_T (p^2)$}.
Therefore we can recognize that the calculation of this diagram is quite different if the technipion wave function (or the vertex) is hard or soft. A rough evaluation of this diagram within the dynamical perturbation theory approach~\cite{Pagels:1979hd} gives
\be
m_\Pi^2 \propto g_A^2 \left( \frac{\mu_{TC}^2}{F_\Pi^2}\right) M_A^2 .
\label{ampi}
\ee
This calculation is certainly very model dependent. However, when the pseudoscalar vertex is soft the result turns out to be suppressed by the $M_A$ mass, and does not increase
with $M_A$ as shown in Eq.~(\ref{ampi}). As a consequence pseudo-Goldstone boson masses turn out to be heavier in the scenario presented here.

Finally a last consequence that results from the slowly decreasing TC self-energy is the possible explanation of why the observed Higgs
boson mass ($m_H$) may be so light compared with the composition mass scale ($\mu_{\tt{TC}}$). The conventional prediction for the
scalar mass in composite models is  $m^2_H \approx 4 \mu^2_{\tt{TC}}$, derived from the homogeneous
Bethe-Salpeter equation (BSE). However, this relation is modified by the inhomogeneous BSE normalization condition when the scalar wave function, which is directly related to the fermion self-energy, is a slowly decreasing function with the momentum. In this case, the above
mass relation is then replaced by $m^2_H \approx 4 \mu^2_{\tt{TC}}f(\alpha_{\T},C_{\T})$ where the function $f(\alpha_{\T},C_{\T})$
depends on the TC gauge group, its coupling constant and fermionic representation, and can be naturally a factor 
of {\cal{O}}$(1/10)$~\cite{us3,us4,us31}!

In this work we have given evidences that radiative corrections to TC(QCD) change the UV technifermion(quark) self-energy behavior. This happens when TC and QCD are embedded in a unified theory as in the Farhi-Susskind model. This fact can be observed by noticing that there are perturbative contributions to the SDE that introduce effective four-fermion interactions like the
one shown in Eq.~(\ref{eq2a}), or by modifications of the gap equation boundary conditions that produce a large mass
anomalous dimension. However the most clear evidence of a hard asymptotic self-energy is shown in Fig.~2, which indicates
generated current masses weakly dependent on the ETC boson mass. As a consequence a large splitting between the masses of different
generations must appear by the imposition of a family or horizontal symmetry in such a way
that TC gives masses preferentially to the third family and QCD to the first one. Of course, the horizontal or family symmetry also generates
an enormous complexity to the coupled set of gap equations, which is peculiar to the attempt of obtaining the fermion mass
spectra in the context of dynamical symmetry breaking, but the origin of these mass spectra is just a consequence of the
two strong interactions aligning with the horizontal symmetry. The experimental consequences of this scenario are that 
the new interactions with TC, which can be simply based on one $SU(2)_{\tt{TC}}$ theory,  mostly occur with the third
fermionic generation; and the scalar
boson mass, the one playing the role of the Higgs boson, can also be lighter than usually thought. The scalar boson mass ($m_H$) relation
with $\mu_{\tt{TC}}$ is modified by the BSE normalization condition, which suppresses the ordinary expected value $m^2_H \approx 4 \mu^2_{\tt{TC}}$.
In this type of model it is possible that TC pseudo-Goldstone boson masses may also turn out
to be large and consistent with the present experimental limits~\cite{us2003,nat}, and all these aspects are under study~\cite{us2}.

\section*{ACKNOWLEDGMENTS}
This research  was  partially supported by the Conselho Nacional de Desenvolvimento Cient\'{\i}fico e Tecnol\'ogico (CNPq)
under Grants No. 305815/2015 (A.C.A.), No. 302663/2016-9 (A.D.), and No.302884/2014 (A.A.N) and  by Grants
No. 2013/22079-8 (A.A.N.), No. 2017/07595, and No. 2017/05685-2 (A.C.A.) of Funda\c{c}\~{a}o de Amparo \`{a} Pesquisa do Estado de S\~ao Paulo (FAPESP).

\end{document}